\documentclass[twocolumn,aps,prb]{revtex4} 

\usepackage{graphics,amsmath,graphicx} 
\newcommand{\cbar}{\bar{c}}
\newcommand{\trph}{{\rm tr}_{(ph)}}
\newcommand{\trc}{{\rm tr}_{(c)}}
\newcommand{\trcbar}{{\rm tr}_{(\cbar)}}
\newcommand{\beq}{\begin{equation}}
\newcommand{\eeq}{\end{equation}}
\newcommand{\beqa}{\begin{eqnarray}}
\newcommand{\eeqa}{\end{eqnarray}}
\newcommand{\Ri}{{\rm {\bf R}}_i}
\newcommand{\Rj}{{\rm {\bf R}}_j}

\begin{document}
\title{Hopping dynamics of interacting polarons}
\author{S. Ciuchi$^{1,2}$ and  S. Fratini$^{3,4}$} 

\affiliation{$^1$ Dipartimento di Fisica and CNISM, 
Universit\`a dell'Aquila, 
via Vetoio, I-67010 L'Aquila, Italy \\ 
$^2$SMC Research Center, INFM-CNR, Roma, Italy\\
$^3$Institut N\'eel - CNRS \& Universit\'e Joseph Fourier, 
BP 166, F-38042 Grenoble Cedex 9, France\\
$^4$Instituto de Ciencia de Materiales de Madrid - CSIC, Sor Juana In\'es de la Cruz 3, E-28049 Madrid, Spain} 

\begin{abstract} 
We derive an effective cluster model to address the transport properties 
of mutually interacting small polarons.
We propose a decoupling scheme where the hopping 
dynamics of any given  particle
is determined  by separating out explicitly the 
degrees of  freedom of its environment, which are treated at
a statistical level. 
The general cavity method developed here shows that 
the long-range Coulomb repulsion
between the carriers leads to a 
net increase of the thermal activation barrier for electrical transport,
 and hence to a sizable reduction of the carrier mobility. 
A mean-field calculation of this effect is provided, 
based on the  known correlation functions of the interacting liquid 
in two and three dimensions. 
The present theory gives a natural explanation of recent experiments
performed in organic field-effect transistors with highly polarizable
gate dielectrics, 
and might well  find application in other classes of polaronic systems
such as doped transition-metal oxides.
\end{abstract} 
\date{\today} 
\maketitle

\section{Introduction}

In recent years, the development of organic electronics has 
triggered
a strong effort towards the understanding of charge transport in   
organic field-effect transistors (OFETs).\cite{RevModPhys} In such devices, the
carriers induced by a gate potential move at
the interface  between  an organic semiconductor and a dielectric.
Unlike their inorganic counterparts, such as Si
MOSFETs, the transport properties in OFETs
are dominated
by the weak transfer integrals between 
the molecular constituents of the organic material: 
the Van der Waals 
inter-molecular bonding leads even in pure
crystalline samples to extremely narrow electronic bands, 
making such systems very sensitive to
interactions.

Recently, a systematic study of rubrene-based 
single-crystalline OFETs fabricated
using gate materials of increasing dielectric polarizability has
revealed that, in the case of  high-$\kappa$ dielectrics, 
the dominant limiting mechanism of electron transport
originates from the coupling with the polar phonons at the
organic-dielectric interface.\cite{natmat} 
This phenomenon, which in wide-band
inorganic semiconductors\cite{WangMahan,HessVogl,MoriAndo,Fischetti,IEEE} 
and in graphene\cite{FratGraphene} only leads to minor
modifications of the electron
mobility,  
can be so effective in organic semiconductors that
it leads to polaronic
self-localization of the carriers on the scale of one or  few molecules.
As a consequence, the  mobility is strongly suppressed and
becomes thermally activated, being due to the incoherent hopping
of small polarons on the molecular lattice.

Because of the increased capacitance of the devices, 
the use of high-$\kappa$ dielectrics also has a second interesting
consequence, as it allows the injection of sufficiently large charge
densities, such that the electrons can no longer be considered as
non-interacting carriers.  
\cite{NJP} Indeed, 
concentrations of the order of $0.1$ carriers/molecule and above have
been reached in rubrene  devices using Ta$_2$O$_5$ as a gate material 
(dielectric constant $\epsilon_s=25$).  
The current-voltage characteristics of such devices 
exhibit strong deviations from linearity 
that cannot be explained in terms of
independent carriers, and have been ascribed to the onset of
electron-electron interactions. \cite{NJP}

The aim of this work is to establish a theory 
for the density-dependent transport properties of mutually interacting
small polarons in the hopping regime.
Although the present derivation is motivated by the physics of
organic-dielectric interfaces, the problem itself is sufficiently
general to find application in other polaronic systems such as
transition-metal oxides\cite{AustinMott} and possibly 
oxide-oxide interfaces \cite{oxide-interf}  
and organic  charge transfer interfaces.\cite{MorpNatMat}
As will become clear in the following, however, OFETs are ideal systems for the
observation of the many-body effects 
studied here, for two reasons. 
First,  the effect of interactions on the polaronic hopping rates 
can be comparatively 
large in organic semiconductors, where
small polarons can exist with activation energies that are 
generally smaller than in 
oxides. 
Secondly, and most importantly, in such devices 
the carrier concentration can be varied 
accurately  by tuning the
gate voltage without the need of chemical
substitution,\cite{RevModPhys,Ahn} 
thus providing a reliable and  unambiguous  
procedure to disentangle  many-body effects from the
properties of individual polarons.

The present paper provides a  detailed derivation of the
theory that was 
used in  Ref.\onlinecite{NJP} to explain the current-voltage
characteristics of rubrene-based high-$\kappa$ OFETs, as well as its
generalization to three-dimensional systems.
We start by introducing an effective cluster model that allows to separate the 
dynamics of any given electron 
from the
remaining carriers in the system,  which are treated as
an external statistical environment. 
The mobility is then obtained by assuming a succession of 
incoherent hopping events,
from a suitable statistical average 
of the inter-molecular hopping rates.
The  theory is
applied to the calculation of the density-dependent transport properties
of small polarons interacting through long-range Coulomb
potentials. 
It is shown that the mutual interactions between carriers
gives rise to a net increase of the polaronic thermal activation
barrier, and consequently
to a sizable  reduction of the mobility.
A mean-field calculation is performed,  providing a
closed-form 
analytical expression for the mobility based on the known pair distribution
function of the interacting  liquid.

The paper is organized as follows. In Sec. \ref{model} we carry out
explicitly the separation between the cluster degrees of freedom and
those of the environment. In Sec. \ref{twositecluster} we focus specifically
on a two-site cluster, which is appropriate in the small polaron
limit. We derive a general formula for the hopping mobility of
interacting small polarons 
and provide a simplified expression describing carrier-carrier correlations 
in the mean-field approximation. In
Sec. \ref{Coulomb} the theory is applied to the problem of polarons
interacting via the long-range Coulomb repulsion.
The results are
discussed in relation to the transport properties of OFETs and of other
known polaronic systems in Sec. 
\ref{conclusion}.

\section{Effective cluster model}

\label{model}

Let us consider a system of electrons moving on a lattice (defined by
the lattice vectors $\Ri$) which  interact with each other and
with the lattice vibrations, as described by the following  Hamiltonian:
\beq
H=H_t+H_{ph}+H_{int}+H_{e-ph}.
\label{Model0}
\eeq
The first term 
\begin{equation}
H_{t}=-t\sum_{<ij>} c^\dagger_i c_j
\label{defHt}
\end{equation}
is the tight-binding Hamiltonian for free electrons, with $t$ the
hopping integral and  $c_i$, $c_i^\dagger$ 
the corresponding destruction and creation operators 
for electrons. The second term
\begin{equation}
H_{ph}=\sum_{j}\frac{1}{2}k X^2_j +\sum_{j}\frac{1}{2M} P^2_j 
\label{Hph}
\end{equation}
describes local (dispersionless) 
phonons of frequency $\omega_0=\sqrt{k/M}$, 
where $X_j$ is the local lattice displacement, $k$ the spring constant
and $M$ the mass. The term
\begin{equation}
H_{int}=\frac{1}{2}\sum_{i,j}n_iV_{ij}n_j
\label{Hint}
\end{equation}
is the electron-electron (e-e) interaction
where $n_i=c^{\dagger}_i c_i$ is the electron density operator at site $i$, and
$V_{ij}=V(\Ri-\Rj)$ a generic  density-density interaction potential.
Finally, 
\begin{equation}
H_{e-ph}=\sum_{i,j}n_ig_{ij}X_j
\label{Heph}
\end{equation}
is the electron-phonon (e-ph) interaction
where $g_{ij}=g(\Ri-\Rj)$ is a non-local density-displacement coupling. 
We shall not give here the precise form of  $g_{ij}$ and $V_{ij}$,
which is not needed for the general formalism developed in
the following paragraphs.  
This  will be provided later, when dealing with specific examples.

\subsection{Derivation of the cluster Hamiltonian}
\label{Hcl}
We now derive an effective cluster model neglecting  the second 
term in Eq. (\ref{Hph}), which is valid at temperatures
$T\gtrsim \omega_0$.  The phonon
kinetic energy term will be reintroduced in  Section II B
to treat the polaron hopping dynamics in the semiclassical adiabatic
approximation. 

Our starting point to evaluate the hopping  mobility of the system
described by Eq. (\ref{Model0}) is to
divide the  lattice into a cluster $(c)$ of finite size, 
in which we allow for quantum electron
hopping, and a remaining part $(\cbar)$ in which the dynamics of the
electrons is neglected. 
This separation is enforced by keeping a finite transfer integral $t$ 
only for electrons within the
cluster $(c)$, while setting $t=0$ 
in $(\cbar)$.
It is then natural to rewrite
the model Eq. (\ref{Model0}) 
by singling out 
the terms which explicitly contain electronic  variables in $(c)$,
that we denote as $H_\bullet$:
\begin{equation}
H_{\bullet}=H^{(c)}_t+H^{(c)}_{int}+H^{(c)}_{e-ph}+H^{(c,\cbar)}_{int}
\label{defHbullet}
\end{equation}
so that
\begin{equation}
H=H_{ph}+
H^{(\cbar)}_{int}+
H^{(\cbar)}_{e-ph}+ H_{\bullet}.
\label{clustersep}
\end{equation}
In Eqs. (\ref{defHbullet}) and (\ref{clustersep}) the  labels
$(c)$ and $(\cbar)$  indicate  that the sums
over {\it electronic} variables are restricted respectively
to the cluster or the environment, 
and $(c,\cbar)$ stands for interactions among electrons
belonging to the two different sub-systems. 
It can be noted that all terms 
in Eq. (\ref{clustersep}) commute with each other.

The proposed separation scheme is formally equivalent to the one used by Pardee
and Mahan\cite{Pardee,Mahan76} to describe electrical conduction in 
solid electrolytes. 
The justification in that case follows from
the large masses of the charge carriers, that are mobile ions.
As a result, the collective rearrangement of
the particles in reaction to a given hopping event is much slower than
the hopping process itself, so that the positions of the carriers in
the environment
can effectively be regarded as static variables 
during the local dynamical evolution. 
In our case
this decoupling is justified due to the exponential
suppression of the carriers' hopping rate associated to polaronic
self-localization.

To derive an effective cluster model
 it is useful to introduce the following
reduced density matrix: 
\begin{equation}
\label{rhored}
\rho_r = \frac{1}{Z}\trph  \trcbar e^{-H/k_BT}
\Pi^{(c)}\delta(Y_i-\sum_jg_{ij}X_j)
\end{equation}
where $\Pi^{(c)}$ indicates the product over the cluster electrons, 
the trace symbols are defined as
\beqa
\trcbar \left(..\right)&=&\sum_{n_i,i \in (\cbar)} \left(..\right)
\label{trcbar}\\
\trph \left(..\right)&=&\int \Pi_i dX_i \left(..\right) \label{trph},
\eeqa
and $Z=\trc \trph  \trcbar e^{-H/k_B T}$.
In Eq. (\ref{trcbar}) we trace over all the electronic 
degrees of freedom which do not belong to the cluster. 
The trace over phononic variables in Eq. (\ref{trph}) is performed 
by assigning the value of the phonon-induced external fields
\begin{equation}
\label{eq:constraints}
Y_i=\sum_j g_{ij} X_j\;\;\;\;i\in {\rm (c)}.
\end{equation}
These are the phononic collective variables 
which act on each site of the cluster, through the non-local e-ph
interaction Eq. (\ref{Heph}). The two steps described above are now
explicitly carried out.

\paragraph{Tracing out the phonons.}
The term $H_{\bullet}$ in Eq. (\ref{defHbullet}) 
depends on the phonons only through the variables $\{Y_i\}$.
We can therefore rewrite Eq. (\ref{rhored}) as
\begin{eqnarray}
\rho_r &=& \frac{1}{Z} \trcbar \trph
e^{-(H_{\bullet}(\{Y_i\})+H^{(\cbar)}_{int})/k_BT} \; \times
\label{partfunct}
\\
&& \times \; \Pi^{(c)}_i\delta(Y_i-\sum_j g_{ij} X_j) e^{-
  (H_{ph}+H^{(\cbar)}_{e-ph})/k_BT}
\nonumber
\end{eqnarray}
where with $H_{\bullet}(\{Y_i\})$ we indicate the explicit dependence of
this term on the collective phonon variables.  
The trace over the original phonons $X_i$ in Eq. (\ref{partfunct}) can be
performed by introducing  the integral representation of the $\delta$ function 
\begin{equation}
\delta(Y_i-\sum_j g_{ij} X_j)= \int \frac{d\omega_i}{2\pi} e^{i \omega_i(Y_i-\sum_j g_{ij} X_j)}. 
\label{partfunct1}
\end{equation}
Performing the gaussian integrals over $\lbrace  X_i \rbrace$ 
and over $\lbrace\omega_i\rbrace$ we obtain
\begin{equation}
\rho_r \propto  \trcbar 
e^{-(H_{\bullet}(Y)+H^{(\cbar)}_{int}+{H}^{(\cbar)}_{eff})/k_BT},
\label{partfunct3}
\end{equation}
where ${H}_{eff}^{\cbar}$ represents 
the effective Hamiltonian resulting from 
the trace over phonons. It can be expressed 
as
\begin{eqnarray}
{H}_{eff}^{(\cbar )}&=& -\frac{1}{2}\sum^{(\cbar)}_{i,j}n_i D_{ij} n_j + 
{{H}^\prime}_{eff}^{(\cbar )} \label{Heff}
\end{eqnarray}
with
\begin{eqnarray}
D_{ij}&=&\frac{1}{k}[g^{2}]_{ij}-\frac{1}{k}\sum^{(c)}_{l,k}  [g^{2}]_{il} [g^{-2}_c]_{lk}  [g^{2}]_{kj}\label{D}\\
{{H}^\prime}_{eff}^{(\cbar )}&=& \frac{k}{2}\sum^{(c)}_{i,j} [g^{-2}_c]_{ij}Y_i Y_j
+\sum^{(c)}_i\sum^{(\cbar)}_j G_{ij}Y_i n_j\label{Hbareff}\\
G_{ij}&=&\sum^{(c)}_l [g^{-2}_c]_{il}[g^{2}]_{lj}\label{G}.
\end{eqnarray}
In the above equations we have introduced the symbol $[g^{-2}_c]$ to denote the inverse
of the matrix $[g^{2}]$ in the cluster sub-space.
As can be seen from Eq. (\ref{Heff}), integrating out the phonon
variables has led to an effective attraction
$D_{ij}$ between the  $(\cbar)$ electrons, whose form is 
given by Eq. (\ref{D}). 
Similarly, 
equations (\ref{Hbareff}) and  (\ref{G}) describe
the effective interactions arising between the
$(\cbar)$ electrons and the collective variables $Y_i$.

 Adding the phonon-mediated interaction  of Eq. (\ref{Heff}) 
to the  bare electron-electron
term $H^{(\cbar)}_{int}$ in Eq. (\ref{clustersep}) yields 
the following screened interaction between the environment electrons
\begin{equation}
\label{eq:Hbar}
\tilde{H}^{(\cbar)}_{int}=\frac{1}{2}\sum_{i,j}n_i \left(
  V_{ij}-D_{ij} \right)
n_j.
\end{equation}
The reduced density matrix can be finally expressed as
\begin{equation}
\rho_r \propto  \trcbar 
e^{-(H_{\bullet}(Y)+\tilde{H}^{(\cbar)}_{int}+
{H^\prime}^{(\cbar)}_{eff})/k_BT}.
\label{partfunct3a}
\end{equation}

Before moving on to the integration of the environment electrons
it is useful to comment on the physical meaning of the two different
contributions to the phonon-induced screening in Eq. (\ref{D}). 
The first term, which leads to the effective potential
\begin{equation}
  \label{eq:Vbar}
  \tilde{V}_{ij}=V_{ij}-\frac{[g^2]_{ij}}{k},
\end{equation}
 represents
the ability of the
polarizable medium to partially screen the electron-electron
interaction. For example, 
starting from the bare Coulomb potential $V_{ij}=
e^2/(\epsilon_{\infty} R_{ij})$ and an electron-phonon 
interaction $g_{ij}$ of the  Fr\"ohlich type, 
it is shown in Appendix \ref{effint} that the inclusion this term yields  $\tilde{V}_{ij}=
e^2/(\epsilon_{s} R_{ij})$,
which correctly reproduces the static 
screening response of a bulk polar dielectric.

The second term in Eq. (\ref{D}) is a cavity field 
which arises due to the constraints in
Eq. (\ref{eq:constraints}), because  not all of the phonons have
been integrated out. It can be viewed as the  part of phonon
screening that is missing due to the existence of the cluster.
 Since it involves the product of two matrices
$[g^2]_{il}[g^2]_{kj}$, which decays faster than  the direct screening
$[g^2]_{ij}$ itself, 
this term becomes negligible when the cluster size is smaller than the average 
interparticle distance. This cavity correction can therefore be 
neglected to lowest order
in the electron concentration, although its actual magnitude depends
on the shape of  the electron-phonon interaction $g_{ij}$
(for example, such cavity field 
is clearly absent in the limit of local e-ph interactions, i.e. 
$g_{ij}\propto \delta_{ij}$).
On the other hand, 
if the cluster is enlarged to attain the size of the entire system, the
two terms in Eq. (\ref{D}) exactly cancel, and only the bare 
electron-electron interaction remains.

\paragraph{Tracing out the electronic environment.}
The trace appearing in Eq. (\ref{partfunct3a}) can be formally carried out
 by introducing two classical fields which couple  linearly 
to the cluster variables $Y_i$ and $n_i$, namely:
\begin{eqnarray}
\label{defetaeps}
\eta_i &=& \sum^{(\cbar)}_j G_{ij}n_j \;\;\; ; \;\;\; i\in c  \\
\epsilon_i &=& \sum^{(\cbar)}_j V_{ij}n_j \;\;\; ; \;\;\; i\in c  .\label{defetaeps2}
\end{eqnarray}
Such fields take into account the 
interactions between electrons in
$(\cbar)$ and the cluster degrees of freedom, as 
contained explicitly in ${H^\prime}^{(\cbar)}_{eff}$ 
and in the direct term $H^{(c,\cbar)}_{int}$.
Substituting these definitions into
Eqs.  (\ref{defHbullet}) and (\ref{Hbareff}) and regrouping terms in
Eq. (\ref{partfunct3a})   
one obtains  the following cluster Hamiltonian:
\begin{eqnarray}
\nonumber
H_{cluster}&=&
-t\sum_{<ij>}^{(c)} c^\dagger_i c_j
+\frac{1}{2}\sum_{i,j}^{(c)} n_iV_{ij}n_j+\sum^{(c)}_i n_i\epsilon_i\!\!\\
\label{Hel}
&+&\sum^{(c)}_iY_i(\eta_i+n_i) +\frac{k}{2}\sum^{(c)}_{i,j} [g^{-2}_c]_{ij}Y_i Y_j.
\label{Hcluster}
\end{eqnarray}
Finally, by enforcing the definitions Eqs. (\ref{defetaeps}), 
(\ref{defetaeps2})
through  the appropriate $\delta$
functions, the reduced density matrix of the cluster can be expressed 
as a trace over the classical variables $\eta_i,\epsilon_i$ of the environment 
\begin{equation}
\rho_r =  \int \Pi^{(c)}_i d\eta_i d\epsilon_i  e^{- H_{cluster}/k_BT}
P\left(\lbrace \epsilon_i\rbrace, \lbrace \eta_i \rbrace\right),
\label{partfunct4}
\end{equation}
whose statistical distribution  is
\begin{eqnarray}
P\left(\lbrace \epsilon_i\rbrace, \lbrace \eta_i \rbrace\right)& \propto &
\trcbar e^{-\tilde{H}^{(\cbar)}_{int}/k_BT}
\Pi^{(c)}_i \delta(\eta_i - \sum^{(\cbar)}_j G_{ij}n_j)\times\nonumber\\
&&\times\delta(\epsilon_i - \sum^{(\cbar)}_j V_{ij} n_j).
\label{distrib}
\end{eqnarray}

To summarize, Eqs. (\ref{Hcluster}), (\ref{partfunct4}) and (\ref{distrib}) 
describe
a finite cluster in which electrons mutually interact
via the bare potential $V_{ij}$, 
and are coupled to collective phonon variables $Y_i$. 
The cluster degrees of freedom are also
subject to random fields $\eta_i$ and $\epsilon_i$ arising from the 
environment electrons. Such fields are
distributed, via  Eq. (\ref{distrib}),  
according to the equilibrium distribution 
of classical particles
interacting through the {\it screened}  Hamiltonian
$\tilde{H}^{(\cbar)}_{int}$ defined in Eq. 
(\ref{eq:Hbar}).

\subsection{Ehrenfest dynamics of the cluster model}

Within the adiabatic regime, the carrier motion is constrained to follow 
the slow dynamics of the phonon
coordinates.\cite{AustinMott,Holstein59} 
To determine the polaron mobility it is therefore necessary  to treat 
explicitly the dynamics of the  $X_i$ that was neglected in
the preceding Section. This can be done by introducing the semi-classical
evolution of the lattice degrees of freedom through the following 
Ehrenfest equations
\begin{equation}
  \label{eq:EhrXi}
  M \ddot{X}_i=-kX_i-\sum_j g_{ij} \langle n_j (t) \rangle.
\end{equation}
In the above equation the average of the electronic operators is taken at fixed
${X_i}$.
To change to the cluster variables $Y_i$ we substitute Eq. (\ref{eq:EhrXi})
into Eq. (\ref{eq:constraints}) for
$i\in (c)$, leading to:
\begin{equation}
  \label{eq:EhrYi}
  M \ddot{Y}_i=-kY_i-\sum_j^{(c)} [g_c^2]_{ij} \langle n_j (t) \rangle
  -\sum_j^{(\cbar)} [g^2]_{ij} \langle n_j \rangle,
\end{equation}
where we have made explicit use of the assumption that the environment
electrons do not evolve in time.
Using  Eq. (\ref{defetaeps}) this can be rewritten as 
\begin{equation}
  \label{eq:EhrYi2}
  M \ddot{Y}_i=-kY_i-\sum_j^{(c)} [g_c^2]_{ij} \left\lbrack
\langle n_j (t) \rangle +\eta_j\right \rbrack.
\end{equation}
The collective phonon variables $Y_i$ are therefore subject to an
external force which depends both on the instantaneous electron density
within the cluster and on the environment degrees of freedom through the
fields $\eta_i$.
It is interesting to observe that in the present treatment, 
the frequency of the collective modes is equal to the bare phonon
frequency $\omega_0=\sqrt{k/M}$. 
The above Eq. (\ref{eq:EhrYi2}) can equivalently
be  derived in a Hamiltonian formulation, by adding a kinetic term
$(2M)^{-1} \sum_{ij}^{(c)} [g_c^2]_{ij} \Pi_i\Pi_j$ 
to  Eq. (\ref{Hcluster}), with $\Pi_i$
the momentum conjugate to $Y_i$.

\section{Small polaron limit}
\label{twositecluster}

The actual choice of the cluster size for practical calculations 
is dictated by the polaron properties,
since it should be large enough to accomodate  the
electronic wavefunction involved in the hopping process.
To keep the discussion simple and provide a physically significant
example of the theory presented so far, we now focus specifically on the small
polaron limit, 
where the electronic wavefunction collapses onto a single molecule.
This situation is realized 
in systems with narrow electronic bands, provided that the
electron-phonon coupling is sufficiently strong. To be specific, 
this occurs when the energy 
of a polaron fully localized on a single
molecular site, $E_P=[g^2]_{11}/2k$,   
is larger than approximately half the free electron bandwidth,
in which case a self-localized state becomes energetically more
favorable than an extended wave. 
The proper  cluster in this case   
consists of two molecules --- the initial (filled) site and the final (empty)
site --- and constitutes the basis for 
the theory of small-polaron transport.\cite{AustinMott,Holstein59,LangFirsov}
We shall explicitly consider  situations where 
the formation of bipolaronic states is ruled out by 
the presence of sufficiently strong repulsive 
interactions between the carriers.\cite{bipol1,bipol2}
Apart from this restriction, 
the results obtained in this Section 
concerning the effect of electron-electron interactions will be generally valid
regardless of the physical origin, and particular form, of $g_{ij}$,
the only requirement being that the  polarons are small.

\subsection{Two-site cluster}
It is shown in Appendix
\ref{app:2site}
that for a singly occupied two-site cluster
the Hamiltonian
Eq. (\ref{Hcluster}) reduces to a spin-boson model, where the electronic
degree of freedom plays the role of a pseudo-spin. 
Introducing the notation  $\sigma_z=n_1-n_2$ and
$\sigma_x=c^+_1c_2+c^+_2c_1$ we obtain
\begin{equation}
H_{sb}=-t\sigma_x+\frac{1}{2}kQ^2-\frac{1}{\sqrt{2}}(gQ+\xi)\sigma_z.
\label{Hsb-app}
\end{equation}
The relative electronic occupation $\sigma_z$ is
coupled to the phonons 
through a single ``interaction coordinate''
\beq
Q=\frac{Y_2-Y_1}{\sqrt{2}\ g}+\frac{g}{k}\frac{\eta_2-\eta_1}{\sqrt{2}}.
\label{defQ}
\eeq
The first term in Eq. (\ref{defQ})
is the direct interaction with the collective
phonons, and the second term 
originates from the residual electron-phonon interaction of Eq. (\ref{G}), 
$\eta_1$ and $\eta_2$ being defined by Eq. (\ref{defetaeps}).  
The parameter $g$ is an effective electron-phonon 
coupling for the two-site cluster, defined through
\beq
\label{eq:geff}
g^2=[g^2]_{11}-[g^2]_{12}.
\eeq
The coupling with the environment electrons $(\cbar)$  also occurs via a
single classical variable
\beq
\xi=(\epsilon _2-\epsilon_1) - \frac{g^2}{k} (\eta_2-\eta_1) 
\label{csi}
\eeq
which takes into account the electronic repulsion $\epsilon_i$ on the
two sites, corrected by 
the appropriate  phonon mediated  attractive terms $\eta_i$. Using
Eq. (\ref{eq:Vbar}), this  can be rewritten as
\beq
\xi= \sum^{(\cbar)}_{j}[\tilde{V}_{2,j}-\tilde{V}_{1,j}]n_j.
\label{csieff}
\eeq
Such ``local field'' represents the energy difference between the
two-sites of the cluster 
in the presence of the potentials of the remaining electrons, screened by the 
lattice polarization.

\subsection{Adiabatic hopping}

In order to determine the polaron hopping rate, we now calculate the
evolution of the dynamical variables $\sigma_z$ and $Q$ within the
cluster in the presence of the local field $\xi$, which by assumption 
is fixed during the time of the hopping process.
The electronic variable $\sigma_z$ evolves quantum-mechanically
through Eq. (\ref{Hsb-app}), while 
the phonon collective variable $Q$ is taken to evolve through the
classical Ehrenfest equations\cite{PaganelliCiuchi} 
Eq. (\ref{eq:EhrYi2}), that reduce to
\begin{equation}
M \frac{d^2 Q}{d t^2} = -k Q -\frac{g}{\sqrt{2}}\langle \sigma_z(t)\rangle,
\label{Ehr}
\end{equation}
where the average of the pseudo-spin is taken  at a given configuration
$Q(t)$.

\begin{figure}
  \includegraphics[width=8.5cm]{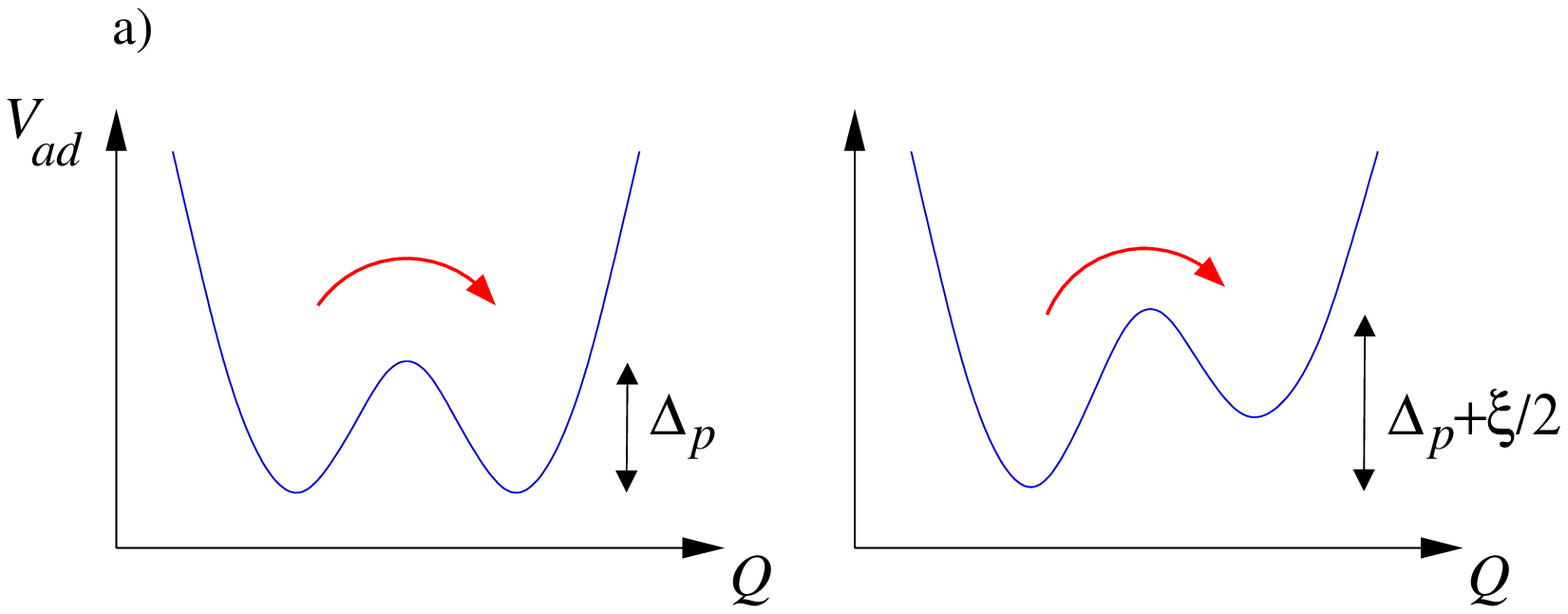}\\
\vspace{0.5cm}
  \includegraphics[width=7.5cm]{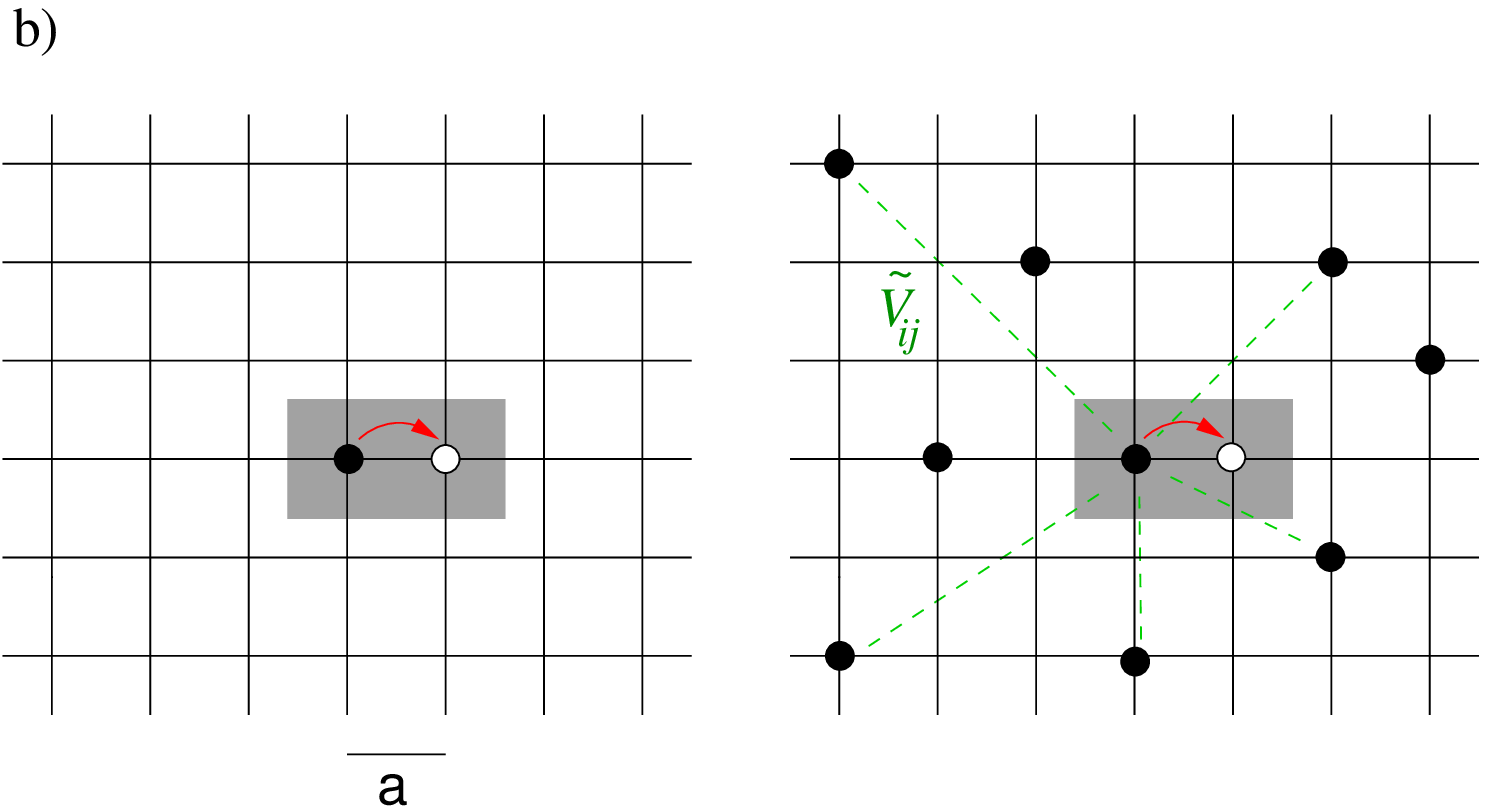}
  \caption{a) Adiabatic double-well potential $V_{ad}(Q)$ of
    Eq. (\ref{Vad-app}) for a 
    two-site cluster, in the absence  (left) and
    in the presence (right) of electron-electron
    interactions.  The electron position follows 
    the dynamics of the phonon coordinate 
    $Q$ between the two minima of the potential. As
    pictorially represented in panel b),  
    interactions between the carriers  
    modify the shape of the double-well via the
    local field   $\xi$, which measures the energy unbalance between the
    two sites due to the presence of the environment electrons. 
     The grid represents the  molecular lattice sites and the shaded
     area is the two-site cluster. The initial and final site of the
     carrier hop are indicated respectively 
    by a black dot and a white circle. }
  \label{fig:2sitepot}
\end{figure}

As a further approximation, 
we estimate the electron transition probability
 within the adiabatic formulation of Refs.
\onlinecite{Holstein59,LangFirsov} 
If the electron dynamics is faster than  the motion of the
phonons, the  quantum variable $\sigma_z$
is able to equilibrate at any given value of the 
classical $Q$.
In this approximation the right-hand side of Eq. (\ref{Ehr}) can be
obtained from the derivative with respect to $Q$ 
of the following adiabatic potential
\beq
V_{ad}(Q)=-k_BT\log {\rm tr}_{\sigma} e^{-H_{sb}(Q,\sigma)/k_BT}.
\label{Vad}
\eeq
At sufficiently 
low temperature [lower than the barrier  $\Delta(\xi)$ defined
below], the
 adiabatic potential reads:
\beq
V_{ad}(Q)=\frac{1}{2}kQ^2-\sqrt{(\xi/\sqrt{2}+gQ)^2/2+t^2},
\label{Vad-app}
\eeq
In the polaronic regime, it has the double-well shape 
illustrated in Fig. \ref{fig:2sitepot}a.
Within the adiabatic description, an electron at site $1$ 
is associated to a 
phononic variable being at the equilibrium point 
$Q\simeq -g/\sqrt{2}k$.
For the occurrence of a classical non-dissipative motion to the
neighboring well, 
the initial kinetic energy of $Q$ 
taken from a Maxwell distribution must exceed the relative maximum of
$V_{ad}$. This defines an energy barrier
\beq
\label{Delta}
\Delta(\xi) = \Delta_P+\frac{\xi}{2} + \frac{k\xi^2}{4g^2}
+\mathcal{O}(t^2),
\eeq
where  $\Delta_P= g^2/4k-t$ is  the activation barrier for 
independent adiabatic polarons.
\footnote{Note that the usually accepted relation stating that 
the barrier $g^2/4k$ for small polarons is half the polaron energy
$E_P=[g^2]_{11}/2k$  only holds for purely local
electron-phonon interactions, $g_{ij}\propto \delta_{ij}$. For longer
range interactions\cite{Mahan}  one has in general
$g^2/4k=\gamma E_P$, with  
$$\gamma=\frac{1}{2}\left( 1-[g^2]_{12}/[g^2]_{11}\right
)<\frac{1}{2}.$$
For the Fr\"ohlich interaction in three dimensions a straightforward
calculation using the Fourier transform of the matrix element 
$M_q\propto 1/q$ on a cubic lattice 
gives $\gamma\simeq 0.3$, while for the electron-phonon
interaction at polar interfaces $\gamma\simeq 1/2 -z/\sqrt{a^2+4z^2}$
monotonically decreases with the distance $z$ to the interface. }
The rate of electron hops per unit time from site $1$ to site $2$  
then acquires a characteristic thermally activated behavior:
\beq
\label{eq:rate}
{\rm w}(\xi) = \frac{\omega_0}{2\pi} \exp[-\Delta (\xi)/k_BT].
\eeq
We note that $\omega_0^{-1}$
is the time it takes for the classical coordinate 
to pass from the initial point at $Q\simeq -g/\sqrt{2}k$ 
to the final point at $Q\simeq 
g/\sqrt{2}k$,  and therefore 
corresponds to the natural timescale of the hopping
process. As can be seen from Eq. (\ref{eq:rate}), in the hopping regime 
the time ${\rm w}(\xi)^{-1}$ between hopping events  is 
exponentially longer than the hopping timescale  $\omega_0^{-1}$.
This fact  allows to
neglect the rearrangement of the environment electrons during a given
hopping process, 
validating the cluster/environment 
separation scheme proposed in the present work.

Finally, we remark that the  adiabatic treatment developed here is 
valid when the condition 
\begin{equation}
  \label{eq:ad}
\frac{\hbar \omega_0}{\pi} \left[ \frac{(g^2/4k)
    k_BT}{\pi}\right]^{1/2} \ll t^2   
\end{equation}
is met.\cite{Holstein59,LangFirsov} In the opposite non-adiabatic
regime, a hopping rate equivalent to the one given by
Eqs. (\ref{Delta}) and (\ref{eq:rate}) is
obtained, although with a different prefactor:
\cite{Schnakenberg,AustinMott,Marcus,Cornil}
\begin{equation}
  \label{eq:hopanti}
{\rm w}(\xi) = \frac{t^2}{\hbar^2}  \left[ \frac{\pi}{4 T \Delta_P}\right]^{1/2} \exp[-\Delta (\xi)/k_BT]
\end{equation}
and with $\Delta_P=(g^2/4k)$.
This has exactly the same dependence as Eq. (\ref{eq:rate}) 
on the local field $\xi$ which
embodies the effects of carrier-carrier interactions.
It therefore appears that the theory developed here for interacting 
small polarons 
holds independently of the adiabatic/non-adiabatic character of the
polaronic transport [Eq.(\ref{eq:ad})], 
provided that the appropriate prefactor is used in the
hopping rate.

\subsection{Small polaron mobility}

To determine the mobility, we assume that the current flow occurs
through a succession of incoherent hopping events.
Each individual process is characterized by a  rate of
the form Eq. (\ref{eq:rate}), which depends explicilty on the electronic 
environment of the hopping particle through its own local field $\xi$.
The calculation of the mobility therefore 
amounts  to averaging the hopping rate 
over all the possible values of the local
field $\xi$ through the appropriate distribution $P(\xi)$. The
mobility can then be written through Einstein's relation as
\begin{equation}
  \label{eq:mob}
  \mu= \frac{e a^2}{k_B T} \langle {\rm w} \rangle
\end{equation}
where $a$ is the length of the electron hop, which we take to be equal to the
inter-molecular distance, and $\langle {\rm
  w}\rangle$ is the statistical average
\beq
\langle {\rm w} \rangle = \int d\xi P(\xi) {\rm w}(\xi).
\label{defw}
\eeq

To find the statistical distribution  that enters in Eq. (\ref{defw})
we observe that if the system is sufficiently close to 
equilibrium,
the value of the (static) local field $\xi$ is determined, 
via Eq. (\ref{csieff}), by the positions of the environment electrons
prior to the hop. Correspondingly, 
$P(\xi)$  follows, via Eq. (\ref{distrib}), 
from the equilibrium distribution of interacting classical   
particles 
{\it constrained to the presence of an electron on the initial
cluster site}.  Such constraint clearly introduces spatial correlations  
between the hopping particle and  the
environment electrons.  
By creating a
``correlation hole'' around each carrier, interactions make
polaron hopping in a finite density liquid 
more unfavorable than for non-interacting
polarons, implying a reduction of the mobility. As will be shown in the next Section, such static correlations are 
reflected in an increase of the activation barrier for
electrical transport.
On the other hand, having implicitly 
assumed that the environment of any given
particle is at equilibrium (i.e. that it relaxes
to equilibrium before the same particle can hop again), 
we are  automatically excluding  {\it dynamic} correlations between subsequent
hops.\cite{Mahan76} 
Preliminary numerical simulations performed by us on the
interacting liquid indicate that such dynamic 
correlations can at most modify the
prefactor of Eq. (\ref{eq:mob}), which amounts to  logarithmic
corrections to the activation barrier. For the present problem of interacting
polarons,  the effect 
would  therefore be negligible 
compared to the effect of spatial correlations that we are actually 
calculating.

Finally, the textbook result\cite{AustinMott} for the 
mobility of independent polarons is recovered by letting $\xi=
0$ in the above equations:
\begin{equation}
  \label{eq:mobind}
  \mu_P= p \frac{e a^2}{ k_B T} e^{-\Delta_P/k_BT }
\end{equation}
with the prefactor
\begin{eqnarray}
  \label{eq:pref}
   p= \frac{ \omega_0}{2\pi} \phantom{xxxxxxxxx} & \mathrm{adiabatic}\\
   p= \frac{t^2}{\hbar^2}  \left[ \frac{\pi}{4 T
       \Delta_P}\right]^{1/2}  &  \mathrm{non-adiabatic}.
\end{eqnarray}

\subsection{Mean-field approximation}

A complete determination of 
the statistical distribution $P(\xi)$ defined in the preceding Section
requires the knowledge of all the many-particle  
correlation functions of the system (generally speaking, the
n-th moment of the distribution is related to an n-particle 
correlation function).
To obtain a tractable expression for the mobility, 
here   we  evaluate the effect of electron-electron interactions
on the average hopping rate Eq. (\ref{defw}) at mean-field level, i.e. neglecting
the fluctuations of the local field $\xi$.
This  scheme of approximation
corresponds to the theory applied in Ref. \cite{NJP} to the study of
organic/dielectric interfaces.  
It amounts to 
substituting the averaged hopping rate Eq. (\ref{defw}) with its first
cumulant
\beq
\label{eq:mfrate}
\langle {\rm w} \rangle \simeq 
p \exp[-\Delta(\langle \xi \rangle)/k_BT].
\eeq
With this replacement, the problem 
can be solved in terms of the sole
{\it two-particle} correlation function of the interacting system,
through the evaluation of the {\it average} local field 
\begin{equation}
  \label{eq:xiavdisc}
  \langle \xi \rangle  =\sum_{j}^{(\cbar)}
  [\tilde{V}_{2,j}-\tilde{V}_{1,j}] \langle n_j\rangle_1,
\end{equation}
where the symbol $\langle n_j\rangle_1$ 
stands for the constrained probability of
occupation of site $j$ with site $1$ occupied. 
As anticipated earlier, while the unconstrained 
average of $\xi$ would clearly vanish by symmetry in a homogeneous system,  
the spatial correlations enforced by this constraint  cause
a net additional energy cost $ \langle \xi
\rangle>0$ for hopping from site to site in the presence of 
repulsive interactions.
Assuming that the interaction correction $\langle
\xi \rangle \lesssim \Delta_P$, so that the quadratic term $\xi^2$
in Eq.   (\ref{Delta}) can be neglected, we obtain a barrier
\begin{equation}
  \label{eq:impcond}
  \Delta(\langle \xi\rangle)=\Delta_P+\frac{\langle \xi\rangle}{2}
\end{equation}
which is the sum of the polaronic activation energy and a many-body correction
term due to interactions. 
From Eq.  (\ref{eq:mfrate})
the density dependent mobility can finally be expressed in terms 
of the mobility of independent polarons Eq. (\ref{eq:mobind}) as
\begin{equation}
  \label{eq:densdep}
  \mu=\mu_P \exp[-\langle \xi \rangle/2 k_BT].
\end{equation}
This result shows that in the regime  $\langle
\xi \rangle \lesssim \Delta_P$
the many-body effects on the mobility are completely
decoupled from the individual polaron properties.

It can be noted that 
Eqs. (\ref{eq:impcond}) and (\ref{eq:densdep}) are formally equivalent to the  
formulas
commonly used to describe impurity conduction 
in compensated polar semiconductors 
and in transition metal oxide glasses.
\cite{AustinMott,Schnakenberg,Emin92,Murawski} 
In such disordered systems, however, the 
microscopic mechanism responsible for the
increase of the polaronic barrier is {\it extrinsic} to the polaronic
system, as it originates from the ability of the particles to find an efficient
percolating path connecting dilute, randomly distributed, impurities.
\cite{Miller-Abrahams}  That 
picture is fundamentally different from the one considered here, where
 $\langle \xi \rangle$ originates
from  the mutual interactions between carriers in a
perfectly crystalline material.

\section{Long-range Coulomb interactions}
\label{Coulomb}

We now apply the theory developed so far
to the calculation of the mobility of a liquid of 
small polarons in the presence of
Coulomb interactions. We shall treat separately the cases of
interacting polarons in two and three space dimensions: the former
applies to the problem of polar interfaces as can be found in
 OFETs with highly polarizable gate dielectrics, while the
latter can be relevant for doped polar semiconductors and oxides with
strong electron-phonon interactions. 
In both situations,
the hopping motion associated to the polaronic nature 
of the charge carriers prevents a proper
screening of the interactions, so that the full long-ranged Coulomb
potential needs to be considered. 
We shall therefore take the general form 
\begin{equation}
  \label{eq:Vtildecoul}
  \tilde{V}_{ij}=\frac{(e^*)^2}{R_{ij}}
\end{equation}
where the effective charge $e^*$ accounts for the dielectric screening
of the polar medium. It is shown in Appendix \ref{effint} that
$e^*=e\sqrt{2/(\kappa+\epsilon_s)}$ at a two-dimensional polar
interface, and $e^*=e/\sqrt{\epsilon_s}$ in a bulk polar material.

We start from the observation that for a Coulomb system 
the correlation 
function $\langle n_j\rangle_1$ appearing in Eq. (\ref{eq:xiavdisc}) varies  on
lengthscales set by the average inter-particle distance $\sim R_s$,
defined as $ R_s=(\pi n)^{-1/2} $ in two dimensions and $ R_s=(4\pi n/3)^{-1/3} $ in three dimensions, $n$ being the particle density. At
sufficiently low concentrations, $R_s$ is much larger than the lattice
spacing so that this function 
can be safely replaced by its continuous limit.
Correspondingly,
the discrete sum appearing in
Eq. (\ref{eq:xiavdisc}) can be replaced by the following integral
\begin{equation}
  \label{eq:xiavg}
  \langle\xi \rangle= n \int d {\bf r} \; [\tilde{V}({\bf r}+
  {\bf R}_{12})-\tilde{V}({\bf r})] \; g^{(2)}(r),
\end{equation}
with $R_{12}=a$ and  $g^{(2)}(r)$ the  
pair distribution function of a classical liquid of interacting
charged particles --- the  one component
plasma (OCP). \cite{Hansen}
The properties of the OCP are 
governed by a single dimensionless coupling parameter 
\begin{equation}
  \label{eq:gammagen}
\Gamma=\frac{(e^*)^2/R_s}{k_B T} 
\end{equation}
measuring the ratio between the
electrostatic interactions and the thermal energy. 
This parameter identifies a weakly correlated and a strongly
correlated regime respectively for $\Gamma\ll 1$ and $\Gamma\gg 1$.
Upon expanding the term between brackets in
Eq. (\ref{eq:xiavg}) to second order in $y=a/R_s$,
it is readily shown that 
$\langle \xi \rangle$
can be expressed in terms of the dimensionless quantities $\Gamma$ and
$y$ 
as
\begin{equation}
  \label{eq:xiavgscal}
\langle\xi \rangle= \frac{k_B T}{2} y^2
  F(\Gamma),
  \end{equation}
with $F(\Gamma)$ a universal function of the OCP. The many-body
effects on the activation barrier are therefore entirely controlled by
the parameter $\Gamma$ characterizing the interacting liquid.

\subsection{2D}
\label{2D}

For a homogeneous two-dimensional 
system,
performing the angular integration in
Eq. (\ref{eq:xiavg}) and integrating the resulting
expression by parts we obtain 
\begin{equation}
  \label{eq:fgamma}
  F(\Gamma)=\Gamma \int_0^\infty dy \frac{g^{(2)}(y)}{y^2} .
\end{equation}

\begin{figure}
  \centering
     \includegraphics[width=8cm]{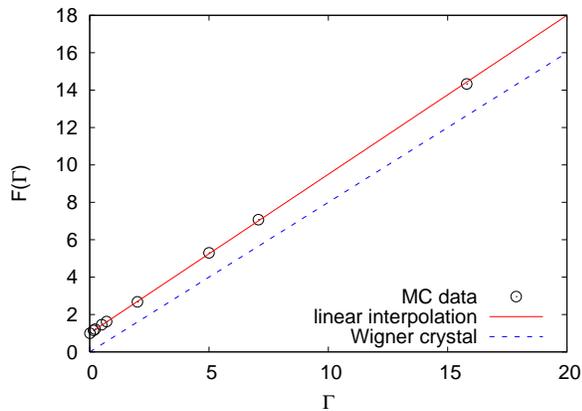}
  \caption{The function $F(\Gamma)$ for a classical two-dimensional
    Coulomb liquid: 
    Monte-Carlo data of Ref.\onlinecite{MC} (open circles),
    interpolating function Eq. (\ref{eq:fgammaMC}) (full line) and 
    Wigner crystal result (dashed line).} 
  \label{fig:MC}
\end{figure}

In the low density/weakly interacting regime $\Gamma \ll 1$, 
the correlations of the classical OCP
are fully determined by the Debye-H\"uckel form 
\cite{Ichimaru}
\begin{equation}
  \label{eq:Debye}
  g^{(2)}(r)=e^{-\tilde{V}(r)/k_B T}=e^{-\Gamma/y}.
\end{equation}
Upon substituting this function into Eq. (\ref{eq:fgamma}) one
obtains $F(\Gamma)=1$.

In the opposite limit of strong coupling, the electronic system
undergoes Wigner crystallization, which occurs for $\Gamma>125$.
\cite{Ichimaru} 
In this regime, it is easy
to calculate the energy corresponding to a spatial displacement $u$ of a
given electron while the remaining particles are kept at rest. Since
the  electron under study is initially in an equilibrium position, the
energy variation is quadratic  in the displacement and can be written
as
\begin{equation}
  \label{eq:WCdev}
  E(u)-E(0)= \zeta
  \frac{(e^*)^2}{2R_s^3} u^2.
\end{equation}
Substituting
$u=a$ and converting into the proper units we obtain
$F(\Gamma)=\zeta \Gamma.$ The value of the numerical constant
$\zeta=0.8$  has been obtained through direct Ewald
summation of the Coulomb interactions on a triangular lattice,\cite{JPCS} 
which is the lowest energy structure of a Wigner crystal in two dimensions.

For the evaluation of $\langle \xi \rangle$
at intermediate interaction strengths  we 
resort to the Monte-Carlo simulations of the classical
two-dimensional OCP performed in Ref.\onlinecite{MC}. 
There the pair distribution function $g^{(2)}(r)$ was
tabulated at different
values of the Coulomb interaction parameter. Upon performing the
integral  Eq. (\ref{eq:fgamma}) using such numerical data, 
one obtains a discrete set of points for the function $F(\Gamma)$.
In the range $1<\Gamma<20$,  
the result can be
parametrized through the linear interpolating function
\begin{equation}
  \label{eq:fgammaMC}
F(\Gamma)=1+0.85 \Gamma  
\end{equation}
within $1\%$ accuracy (cf. Fig. \ref{fig:MC}), and  this formula
remains fairly accurate even at larger values of $\Gamma$, until it
eventually merges into the strong coupling Wigner crystal estimate. 
It can be observed that, except for a constant preasymptotic
term of order $1$, the function $F(\Gamma)$ representing the
interparticle correlations in the Wigner crystal  has essentially the
same $\Gamma$ dependence as that of the correlated liquid.
Using  Eqs. (\ref{eq:gammagen}), (\ref{eq:fgammaMC})  
and  the definition of $R_s$,
we can finally write  
the many-body correction to the polaronic activation barrier due to
Coulomb interactions as
\begin{equation}
  \label{eq:xiavg2Dhuman}
  \langle
    \xi\rangle =\frac{\pi}{2} n a^2   \left[k_BT + 0.85 (e^*)^2 (\pi
      n)^{1/2} \right]. 
\end{equation}
The average local field becomes temperature independent and behaves
asymptotically as $\langle \xi
 \rangle \propto n^{3/2}$ in  the 
 strongly correlated limit ($\Gamma \gg 1$), i.e.
 when the  second term between brackets dominates.

\begin{figure}
  \centering
  \includegraphics[width=8cm]{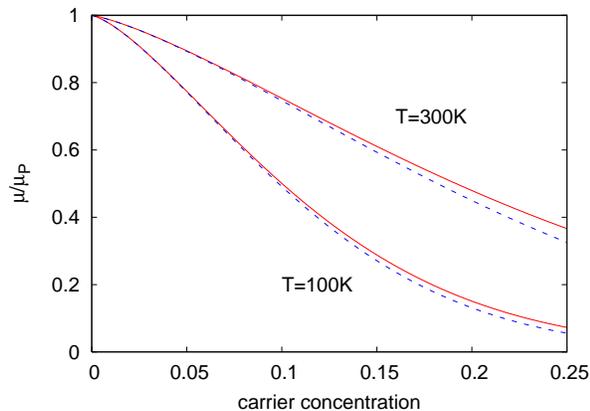}
  \caption{Interaction induced reduction of the polaronic mobility,
    calculated  with the parameters appropriate
    to a rubrene/Ta$_2$O$_5$ interface, 
at two different temperatures.  The full curves are
    obtained from Eq. (\ref{eq:densdep}), while the dashed curves include
  the full activation barrier  given by Eq. (\ref{Delta}), with
  $\Delta_P=55meV$. The differences are negligible because in all the
  explored range $\langle\xi\rangle \lesssim \Delta_P$. Similar
  curves are obtained  for bulk 
  transition-metal oxides.}
  \label{fig:densdep}
\end{figure}

Considering the effective polaron-polaron interaction derived in  
Appendix \ref{effint} for  organic/dielectric interfaces, 
and using the parameters appropriate to a rubrene/Ta$_2$O$_5$ devices
of Ref. \onlinecite{NJP}  ($a=7.2\AA$, $\epsilon_s=25$,
$\kappa=3$), we infer that a moderately correlated polaron
liquid is realized in the two-dimensional conducting channel, 
with coupling parameters in the range $0 < \Gamma
\lesssim 9$.  In this regime  polaron-polaron correlations 
 yield an  increase of the
activation barrier for transport reaching  $\langle \xi \rangle/2
\sim 5 meV$ at the highest concentrations  measured. 
This is  smaller than the barrier $\Delta_P=55 meV$ 
for independent polarons, consistent with the
assumptions underlying our derivation. 
When substituted into Eq. (\ref{eq:densdep}),  such many-body correction 
leads to a sizable  reduction of the polaronic mobility,  
as illustrated in 
Fig. \ref{fig:densdep} at two different temperatures.
\footnote{It can also be verified that the condition for adiabatic transport 
 is fulfilled in such devices, as the ratio between the l.h.s. and the
 r.h.s. of Eq. (\ref{eq:ad}) is $\sim 0.3$ at room temperature.
Actually such value places these devices
close to the adiabatic/nonadiabatic crossover. In this regime, the
polaronic activation energy  changes from $\Delta_P=(g^2/4k)-t$ to the
non-adiabatic value $\Delta_P=(g^2/4k)$, which could be at the
origin of the missing correction $-t$ 
in the activation energy reported in Refs. \onlinecite{natmat,NJP}.}

\subsection{3D}

In three space dimensions,
\begin{equation}
  \label{eq:fgamma3D}
  F(\Gamma)=\Gamma \int_0^\infty dy \frac{d}{dy} g^{(2)}(y)=\Gamma. 
\end{equation}
This result, which follows directly from the fact that in a homogeneous liquid
phase $g^{(2)}(\infty)=1$
(the pair correlations vanish at large distances),  
holds {\it exactly} at all $\Gamma$. 
It is therefore not necessary to integrate numerically the pair distribution
function obtained from Monte-Carlo simulations as was done in the
two-dimensional case. 
It can be directly checked that the result Eq. (\ref{eq:fgamma3D}) 
also extends to the crystallized phase. To this aim we  observe that  
the energy cost to
displace a particle from its equilibrium position in a
three-dimensional Wigner crystal 
is  still given by Eq. (\ref{eq:WCdev}), with now $\zeta^{(3D)}=1$ from  
Gauss' theorem,\cite{Mahan}  also leading to
$F(\Gamma)=\Gamma$.

For practical calculations the result can be rewritten in terms of the
carrier density $n$ in a generic three-dimensional system as
\begin{equation}
\label{eq:xiavg3Dhuman}
  \langle \xi \rangle =\frac{2\pi}{3} \frac{e^2}{\epsilon_s} na^2,
\end{equation}
which is obtained by substituting
Eq. (\ref{eq:fgamma3D})
into Eq. (\ref{eq:xiavgscal}).
As usual,  $a$ is the  hopping distance, of the order of the lattice spacing.
The linear density dependence of the local field  resulting from
Eq. (\ref{eq:xiavg3Dhuman}) is  
 weaker than the $n^{3/2}$ behavior obtained in two dimensions, and is
temperature independent at all densities. 

Using typical values for
transition-metal oxides such as $a=4\AA$ and 
$\epsilon_s=10-100$ and assuming a cubic lattice structure for simplicity
we obtain a barrier increase $\langle \xi \rangle/2 =\alpha x$, where
$x$ is the carrier concentration and  
the coefficient $\alpha\simeq  40-400 meV$. 
An increase of activation energy with
electron concentration compatible with such prediction  has been observed 
in doped three-dimensional transition metal oxides exhibiting small polaron
conduction, such as  magnetite \cite{Kozlowski} 
and the  manganites \cite{Worledge,Palstra}. 
Actually, in the high temperature phases of the manganite  compounds 
 La$_x$Ca$_{1-x}$MnO$_3$, 
both a large polaron scenario (in bulk
samples\cite{Neumeier05}) and a small polaron scenario (in thin
films\cite{Worledge}) have   have been invoked to interpret the
transport properties in the lightly electron-doped regime.
We have performed a linear fit of the
doping dependence of the activation energy $\Delta$ reported in 
Ref. \onlinecite{Worledge} in the range
$0<x<0.35$, yielding  $\Delta=46+56x \; meV$. When compared with
Eqs. (\ref{eq:impcond}) and 
(\ref{eq:xiavg3Dhuman}),  the fitted slope of the
concentration-dependent term yields  $\epsilon_s\simeq 70$, in good
agreement with the dielectric constants measured in those compounds
($\epsilon_s\simeq 55-90$ from Ref. \onlinecite{Cohn}).

\begin{table}
  \centering
  \begin{tabular}{|c|c|c|}
    \hline
    2D & $R_s=(\pi n)^{-1/2} $ &$\displaystyle \langle
    \xi\rangle =\frac{\pi}{2} n a^2   \left[k_BT + 0.85 (e^*)^2 (\pi n)^{1/2} \right]$\\
\hline
    3D & $R_s=(4\pi n/3)^{-1/3}$& $\displaystyle \langle
    \xi\rangle=\frac{2\pi}{3}(e^*)^2 n a^2 $  
\\ \hline 
  \end{tabular}
  \caption{Summary of the main formulas determining the density-dependent
    mobility of small polarons interacting through the long-range
    Coulomb potential of Eq.(\ref{eq:Vtildecoul}), in two and three
    dimensions.  $R_s$ is the mean
    interparticle separation,  $a$ the hopping distance, equal to
    the distance between molecular units, an $e^*$ is the effective
    charge determined by the 
    dielectric environment (see Appendix B).  The right column
    follows from Eqs.  (\ref{eq:gammagen}), (\ref{eq:xiavgscal}),
    (\ref{eq:fgamma}) and (\ref{eq:fgamma3D}). It
gives the mean-field correction to the polaron mobility due to
many-body effects through  $\mu/\mu_P=\exp(-\langle \xi\rangle /2k_BT)$ 
       [Eq.(\ref{eq:densdep})].  
} 
  \label{tab:summary}
\end{table}

\section{Discussion and conclusions}
\label{conclusion}

In this work we have derived a theory for the hopping
transport of mutually interacting polarons in narrow band
materials. Observing that in the hopping regime 
the quantum coherence of the carriers extends
over only few lattice sites, we  solve for the quantum dynamics of the
carriers within  a finite size cluster, taking into account the
interactions with the other charges in the environment via a set of
static fields.
The calculation 
then proceeds by assuming that transport occurs
through statistically independent hopping events. 
Correspondingly, the many-particle mobility is  obtained
from a 
statistical average of the inter-molecular hopping rates over the 
distribution of environment fields, which follows from the
known statistical properties of the interacting liquid.  

The proposed decoupling scheme, which
is analogous to the one followed by Pardee and Mahan\cite{Pardee,Mahan76} in
the context of ionic conductors, is justified here by
the quasi-static nature of the carriers in the hopping regime
as a consequence of polaronic 
self-trapping. 
Despite this simplification, 
which amounts to neglecting dynamical
correlations between subsequent hops, the spatial correlations between
particles  which constitute the dominant many-body effects on
polaronic transport are fully retained. 
When applied to
a liquid of small polarons interacting through 
long-range Coulomb forces, 
the theory predicts 
a net increase of the activation barrier for
electrical transport and hence a reduction of the carrier mobility.
The analytical formulas obtained at mean-field level, i.e. neglecting
the fluctuations of the environment field $\xi$ representing the
polaron-polaron correlations, are 
summarized in Table \ref{tab:summary}.

The present scenario consistently explains the current characteristics 
of rubrene/Ta$_2$O$_5$ OFETs measured in Ref.  \onlinecite{NJP}. 
There, a saturation of the usual linear $I \propto V_g$
relationship expected for independent carriers 
was observed at large values of the gate voltage $V_g$,
indicative of a sizable reduction of the 
mobility (the reader is referred to that work for a detailed
comparison with the experimental data).
An interpretation in terms of 
carrier-carrier interactions  comes naturally 
in these devices where, as was mentioned in the introduction, 
all  the conditions for the 
observation of the predicted many-body effects on
the polaronic hopping transport are  simultaneously met:
small polaron formation 
(because of the strong polar coupling with the gate dielectric and the narrow
bandwidth of the organic semiconductor),
long-range
Coulomb repulsion between the carriers and broad tunability of the carrier
concentration via the applied gate potential.  

We  anticipate based on our theoretical results that, in principle,
nothing prevents the observation of a downturn of the $I-V_g$ curves  
beyond the saturation regime observed in Ref. \onlinecite{NJP}. For
this, the only requirement is that of
a stronger reduction of the mobility than the one realized at
rubrene/Ta$_2$O$_5$ interfaces. 
As is clear from Fig. \ref{fig:densdep},
this can be achieved either by reducing the 
temperature, or by increasing the carrier density, as both effects
lead to an increase of the correlation parameter $\Gamma$ [see 
Eq. (\ref{eq:gammagen})] and therefore of the ratio $\langle \xi
\rangle/2k_BT$ in Eq. (\ref{eq:densdep}). 
An interesting possibility in this direction
is offered by the use of polar electrolytes as gate materials,
allowing to reach much higher concentrations than with conventional 
polar dielectrics.\cite{Shimotani,Panzer}

Finally, due to the very general nature of the mechanisms
involved, one might ask if similar effects can be observed in other
classes of systems. In principle, any system with a sufficient
concentration of small polarons (whatever the microscopic origin) 
interacting through long-range repulsive forces
should exhibit a density-dependent increase of the transport
activation energy.
In fact, we have found at least two examples in the literature which
could fit in the present scenario. 
In the manganite compound La$_x$Ca$_{1-x}$MnO$_3$, 
systematic experimental studies of 
polaronic transport 
in both thin films\cite{Worledge} 
and bulk samples \cite{Palstra}  have reported a monotonic 
increase of activation barrier upon increasing the
electron concentration $x$, that could be  
ascribed to polaron-polaron interactions.\cite{Palstra} 
A similar increase has been observed in Ti doped magnetite
(Fe$_{3-y}$Ti$_y$O$_4$)\cite{Kozlowski}, where a possible explanation in
terms of long-range Coulomb interactions between the carriers
has also been explicitly suggested.  
In both classes of compounds, the linear increase of the activation
energy with electron doping
is indeed compatible with the predictions of our theory.
Nevertheless, other mechanisms can not be excluded, related to 
the complex structural details of these materials,
as well as to the presence
of randomly distributed ionized dopants, whose electric fields 
could also affect the
polaronic hopping rates.

We conclude by suggesting an experimental method that could be useful to
disentangle more clearly the effects of polaron-polaron interactions
from the intrinsic features of non-interacting polarons. 
Such method
relies on the comparison of the activation energy $\Delta$ determined from
electrical transport, and $\Delta_S$ obtained from thermoelectric power
measurements. Since the thermopower is insensitive to the polaronic
renormalization of the carriers, $\Delta_S$ would give a direct measure of
the interaction correction $\langle \xi \rangle/2$ alone, 
while electrical
transport would be governed by the sum
$\Delta=\Delta_P+\langle \xi \rangle/2$. 
Such method has been often applied to disentangle impurity effects
from polaron effects in transition metal oxide glasses
\cite{Murawski,Burns} and has also been proposed in the context of ionic
conductors.\cite{Girvin,Mahan76}    
Comparative analysis of the electrical and thermal transport have also
been performed in the  manganite compounds, to ascertain
the polaronic nature of the charge carriers. \cite{Palstra,Jaime}
The feasibility of thermoelectric power measurements
in OFETs has been recently demonstrated in Ref.\onlinecite{Batlogg},
and could provide further
independent insught into the many-body physics of organic field-effect
transistors.

An extension of the present theory to include the effects of
polaron-polaron correlations 
beyond the mean-field approximation, as well as its generalization to
disordered systems, is underway.

\acknowledgments

S.F. acknowledges useful discussions with M.J Calder\'on and
A.F. Morpurgo, and financial support from 
CONSOLIDER CSD2007-0010. 
S.C. acknowledges useful discussions with C. Pierleoni and 
financial support from the
Research Program MIUR-PRIN 2005.

\appendix
\section{ Hamiltonian of a two site cluster}
\label{app:2site}
For a two site cluster, the  Hamiltonian Eq. (\ref{Hcluster}) 
explicitly reads 
\begin{eqnarray}
H_{cl}&=&H_t+V_{1,2}n_1n_2+Y_1\eta_1+Y_2\eta_2+ \label{Hcl2site}
\\
 &+& n_1(\epsilon_1+Y_1)+n_2(\epsilon_2+Y_2)+\nonumber
\\
&+&\frac{k}{2([g^2]_{11}^2-[g^2]_{12}^2)}\left (
  [g^2]_{11}(Y^2_1+Y^2_2)-2[g^2]_{12}Y_1Y_2\right). \nonumber
\end{eqnarray}
For the present problem, we can assume without loss of generality 
that the cluster is singly occupied (there is one electron on the initial
site, the other site being empty for the hopping process to be allowed). 
It is then 
possible to rewrite the cluster Hamiltonian in a form which is formally
equivalent to that of a   tunneling charge
interacting with {\it a single} effective mode, which is essentially a
spin-boson model. 
Defining the couplings
\beqa
\bar{g}^2&=&[g^2]_{11}+[g^2]_{12}\\
g^2&=&[g^2]_{11}-[g^2]_{12},
\eeqa
introducing the  new variables
\beqa
X&=&\frac{Y_1+Y_2}{\sqrt{2\bar{g}}}\nonumber\\
x&=&\frac{Y_2-Y_1}{\sqrt{2g}}\label{YtoX}\\
E&=&\frac{\epsilon_1+\epsilon_2}{\sqrt{2}}\nonumber\\
\epsilon&=&\frac{\epsilon _2-\epsilon_1}{\sqrt{2}}\label{Eepsilon}\\
{\cal N}&=&\frac{\eta_1+\eta_2}{\sqrt{2}}\nonumber\\
\eta&=&\frac{\eta_2-\eta_1}{\sqrt{2}}, \label{Neta}
\eeqa
and enforcing the single occupancy
within the cluster through the condition $n_1+n_2=1$
we can rewrite Eq. (\ref{Hcl2site}) as
\begin{eqnarray}
H_{cl}&=&H_t-\epsilon\frac{n_1-n_2}{\sqrt{2}}+\label{Hcl2sitebis}
-gx(\frac{n_1-n_2}{\sqrt{2}}-\eta)+\phantom{xxxx}\\ 
&&+\frac{E}{\sqrt{2}}+\bar{g}X(\frac{1}{\sqrt{2}}+{\cal N})
+\frac{1}{2}k X^2+\frac{1}{2}k x^2. \nonumber
\end{eqnarray}
From Eq. (\ref{Hcl2sitebis}) we see that 
the variables $\eta$ only contribute to an unimportant shift in the $x$ 
equilibrium position. It is therefore convenient to introduce the deviation 
$Q=x-g\eta/k$ as well as a new interaction variable
\beq
\xi=\sqrt{2}\left(\epsilon-g^2\eta/k\right)
\label{csi-app}
\eeq
which takes into account both the elecrton-electron interaction and
the  electron-phonon screening correction.

Dropping all terms  
which do not couple to the site occupations or 
to the phonon displacement, and introducing
the pseudo-spin notation $\sigma_z=n_1-n_2$, $H_t=-t\sigma_x$, the Hamiltonian 
$H_{cl}$ can finally be written as 
\begin{equation}
H_{sb}=\frac{1}{2}kQ^2+\frac{1}{\sqrt{2}}(gQ+\xi/\sqrt{2})\sigma_z-t\sigma_x.
\label{Hsb}
\end{equation}

\section{Effective electron-electron interactions}

\label{effint}

\subsection{Organic/dielectric interfaces}

In organic
field-effect transistors, charge carriers accumulate
in a two-dimensional layer located at the interface between an organic
crystal and a polar gate dielectric.\cite{RevModPhys} 
The model Eq. (\ref{Model0}) therefore consists of  two-dimensional tight
binding electrons interacting with the polar phonon modes of the interface.
In Fourier space, the electron-phonon interaction matrix element 
has the simple form
\cite{Sak,WangMahan,HessVogl,MoriAndo}
\begin{equation}
  \label{eq:gq}
  M_q=M_0 e^{-qz}/\sqrt{q}
\end{equation}
where  $q$ is the momentum parallel to the interface,   
 $z$ is the distance of the
electrons to the polar interface, which acts as a short-distance
cutoff, and $M_0$ is a coupling constant that depends on the
dielectric properties of the interface.
\footnote{It should be stressed that the interaction Eq. (\ref{eq:gq})
was derived from the 
macroscopic laws of electrostatics, that are valid at distances $>a$. 
In real interfaces, the discrete nature of the polarizable
medium should lead to an additional short-range 
cut-off at lengths of the order of the inter-ionic
spacing. To a first approximation, 
this effect can be incorporated  by treating  $z$  as an effective
phenomenological quantity which 
includes both the channel-interface distance and the lattice
cutoff.}
 It is given by 
$M_0^2=2\pi \hbar \omega_0 e^2 \beta/S$, with $S$
the total surface of the system, and $\omega_0$ the frequency of the
coupled dispersionless polar mode.
The parameter $\beta$ is
a combination of the known dielectric constants of the two
media that constitute the interface, which determines the 
strength of the electron-phonon coupling.
In the present example of an organic/dielectric interface, 
 $\beta=(\epsilon_s-\epsilon_\infty)/(\epsilon_s+\kappa)/
(\epsilon_\infty+\kappa)$ where $\kappa$ is the (frequency
independent) dielectric constant of the organic semiconductor, and
$\epsilon_s,\epsilon_\infty$ are respectively the static and
high-frequency  dielectric constants of the polarizable dielectric.

We start with the ``bare'' interaction potential $V_{ij}$ between 
two charges located at a distance $z$ from the interface:
\begin{equation}
  \label{eq:Vijelec}
  V_{ij}= \frac{e^2}{\kappa}\left\lbrack \frac{1}{R_{ij}} 
-\frac{1}{\sqrt{R_{ij}^2+4z^2}}\frac{\epsilon_\infty-\kappa}
{\epsilon_\infty+\kappa} \right \rbrack,
\end{equation}
where $\epsilon_\infty$   accounts for the high frequency
 electronic polarizability of the
polar material.
To determine the effective potential  we evaluate
\begin{equation}
  \label{eq:gij2cont}
  [g^2]_{ij}/k =\int \frac{d^2q}{(2\pi)^2}e^{-iq R_{ij}} M_q^2/k = 2 \beta e^2 
\frac{1}{\sqrt{R_{ij}^2+4z^2}}
\end{equation}
and with Eq. (\ref{eq:Vbar}) we obtain 
\begin{equation}
  \label{eq:Vijbar}
  \tilde{V}_{ij}= \frac{e^2}{\kappa}\left\lbrack \frac{1}{R_{ij}} 
-\frac{1}{\sqrt{R_{ij}^2+4z^2}}\frac{\epsilon_s-\kappa}
{\epsilon_s+\kappa} \right \rbrack.
\end{equation}
This result is equivalent to what one would obtain from a simple image
charge calculation, considering the full static dielectric constant
$\epsilon_s$ of the polar material right from the beginning.\cite{JPCS}

It was shown in Ref. \onlinecite{natmat} that the conduction
in organic FETs 
effectively takes place within the first molecular layer nearby the
interface. The cut-off distance $z$ is therefore of the order of the
lateral size of the molecules, which is comparable with the 
lattice spacing $a$ itself. At concentrations
such that the typical inter-particle spacing $R_s$ is much larger
than both $a$ and $z$, 
the effective interaction potential  Eq. (\ref{eq:Vijbar})
reduces to
\begin{equation}
  \label{eq:Vijbarlong}
  \tilde{V}_{ij}=\frac{2}{\epsilon_s+\kappa} \frac{e^2}{R_{ij}}.
\end{equation}
which corresponds to a long ranged Coulomb potential with a screened charge 
$e^*=e\sqrt{2/(\epsilon_s+\kappa)}$.

\subsection{Bulk polar materials}

In three-dimensional polar systems one starts with the bare interaction potential
\begin{equation}
  \label{eq:V3D}
  V_{ij}=\frac{e^2}{\epsilon_\infty R_{ij}}
\end{equation}
where $\epsilon_\infty$ accounts for the high frequency polarizability
of the material. The interaction of the electrons with the polar phonon
modes is described by the Fr\"ohlich matrix element $M_q=M_0/q$, with
$M_0^2=2\pi \hbar \omega_0 (e^2/\tilde{\epsilon})/\Omega$. Here $\Omega$
is the total volume of the system,  $\omega_0$ the frequency of the
coupled dispersionless phonon mode and
$\tilde{\epsilon}=(\epsilon_\infty^{-1}-\epsilon_s^{-1})^{-1}$ an
effective dielectric constant. Including the screening effect of the
polar modes as given by Eq. (\ref{D}) correctly yields
\begin{equation}
  \label{eq:V3Dscr}
  \tilde{V}_{ij}=\frac{e^2}{\epsilon_s R_{ij}}
\end{equation}
corresponding to a screened charge $e^*=e/\sqrt{\epsilon_s}$.

\subsection{Local interactions}
To conclude this Appendix we observe that
local electron-phonon interactions as the ones described by the
Holstein model do not give rise to a long-range screening term. This
can be readily seen from Eq. (\ref{eq:Vbar}), where $[g^2]_{ij}/k\propto
\delta_{ij}$. The effective electron-electron interactions are therefore of the
unscreened form $V_{ij}=e^2/\epsilon_{\infty}R_{ij}$ in bulk materials, and
$V_{ij}=2e^2/(\kappa+\epsilon_{\infty})R_{ij}$ at interfaces. 
For a given carrier density, the coupling 
parameter $\Gamma$ is therefore larger than in the case of polar screening, 
and the interaction
effects on the mobility should be correspondingly enhanced.

\end{document}